\documentclass[runningheads]{llncs}

\usepackage[T1]{fontenc}
\usepackage{cite}
\usepackage{amsmath,amssymb,amsfonts}
\usepackage{algorithmic}
\usepackage{graphicx}
\usepackage{textcomp}
\usepackage{xcolor}
\usepackage{booktabs}
\usepackage{comment}
\usepackage{orcidlink}
\usepackage{tikz}
\usepackage[absolute,overlay]{textpos}

\newcommand{\pol}[1][short]{%
  \ifthenelse{\equal{#1}{short}}{PoL}{Proof-of-Location}%
}

\begin{document}

\title{A Taxonomy and Methodology for Proof-of-Location Systems}
\titlerunning{A Taxonomy and Methodology for Proof-of-Location Systems}

\author{
Eduardo Brito\inst{1,3}\orcidlink{0009-0002-9996-6333} \and
Fernando Castillo\inst{2}\orcidlink{0009-0003-6835-8711} \and
Liina Kamm\inst{1}\orcidlink{0000-0003-1479-2195} \and
Amnir Hadachi\inst{3}\orcidlink{0000-0001-9257-3858} \and
Ulrich Norbisrath\inst{3}\orcidlink{0000-0002-6151-991X}}
\authorrunning{Eduardo Brito et al.}
% First names are abbreviated in the running head.
% If there are more than two authors, 'et al.' is used.
%

\institute{Cybernetica AS, Estonia 
M\"aealuse 2/1, 12618 Tallinn, Estonia\\
\email{\{eduardo.brito, liina.kamm\}@cyber.ee} \and TU Berlin, Einsteinufer 17, Germany\\
\email{fc@ise.tu-berlin.de} \and
University of Tartu, Narva mnt 18, 51009, Tartu, Estonia\\
\email{\{amnir.hadachi,ulrich.norbisrath\}@ut.ee}
}

\maketitle              % typeset the header of the contribution
\begin{abstract}
Digital societies increasingly rely on trustworthy proofs of physical presence for services such as supply-chain tracking, e-voting, ride-sharing, and location-based rewards. Yet, traditional localization methods often lack cryptographic guarantees of where and when an entity was present, leaving them vulnerable to spoofing, replay, or collusion attacks. In response, research on Proof-of-Location (\pol{}) has emerged, with recent approaches combining distance bounding, distributed consensus, and privacy-enhancing techniques to enable verifiable, tamper-resistant location claims.

As the design space for \pol{} systems grows in complexity, this paper provides a unified framework to help practitioners navigate diverse application needs. We first propose a taxonomy identifying four core domains: (1) cryptographic guarantees, (2) spatio-temporal synchronization, (3) trust and witness models, and (4) interaction and overhead. Building on this, we introduce a methodology to map application-specific requirements onto appropriate \pol{} architectures. We illustrate this process through three use cases (retail e-coupons, supply chain auditing, and physical e-voting), each showing how different constraints shape protocol choices. Overall, this work offers a structured approach to building secure, scalable, and interoperable \pol{} systems.
\end{abstract}

\begin{tikzpicture}[remember picture, overlay]
  \node[anchor=south, yshift=1cm] at (current page.south) {
    \fbox{%
      \begin{minipage}{0.85\textwidth}
        \footnotesize
        \textbf{Preprint.} This work has been accepted to the 29th International Conference on Enterprise Design, Operations, and Computing (EDOC 2025).
      \end{minipage}
    }
  };
\end{tikzpicture}

\section{Introduction}
\label{sec:introduction}
Location-based services are increasingly central to digital societies—from supply chain management and civic processes to smart city initiatives~\cite{nosouhi2020blockchain,pournaras2020proof,brito2025decentralized}. However, most localization systems (e.g., GPS) and centralized location managers lack cryptographic guarantees of where and when a device was physically present~\cite{waters2003secure,saroiu2009enabling}, leaving them vulnerable to spoofing, replay, and falsification~\cite{dupin2018location,nasrulin2018robust}.

Proof-of-Location (\pol{}) protocols address these gaps by enabling tamper-resistant, verifiable spatio-temporal claims~\cite{amoretti2018blockchain,foam-white-paper}. Modern approaches combine distance bounding, multi-witness attestation, distributed ledgers, and privacy-preserving techniques (e.g., zero-knowledge proofs) to minimize reliance on centralized trust~\cite{bogdanov2025zero,brito2025decentralized,li2020privacy,wu2020blockchain}. Real-world applications span e-voting, supply chain tracking, retail coupons, and digital content authentication~\cite{saroiu2009enabling,brito2025decentralized}.

Yet, no single \pol{} protocol satisfies all use-case demands. Some require sub-meter, real-time guarantees (e.g., e-voting), while others tolerate coarser, ex-post proofs (e.g., location-based marketing). Privacy needs also vary, prompting trade-offs between exposure and overhead. Additionally, deployments involving consensus or multiple witnesses eventually increase complexity~\cite{zhu2011applaus,dupin2018location,brito2025decentralized}. To address these design challenges, this paper makes the following contributions:
\begin{enumerate}
    \item A \textbf{taxonomy} of \pol{} systems, structured around four core domains: (i)~cryptographic guarantees, (ii)~spatio-temporal synchronization, (iii)~trust and witness models, and (iv)~interaction and overhead;
    \item A \textbf{methodology} for mapping application requirements (e.g., precision, adversary model, privacy) to suitable \pol{} architectures;
    \item \textbf{Use-case examples} demonstrating how constraints shape design choices, spanning from retail to civic and logistics contexts.
\end{enumerate}

The rest of the paper is organized as follows. Section~\ref{sec:background} reviews the evolution of \pol{} protocols, from centralized architectures to decentralized and privacy-preserving designs. Section~\ref{sec:taxonomy} introduces our taxonomy of design domains, followed by Section~\ref{sec:methodology}, which outlines a methodology for matching application requirements to protocol features. Section~\ref{sec:usecases} applies this framework to concrete use cases, and Section~\ref{sec:conclusion} closes with conclusion and future directions.

% -- motivate need, outline contributions

\section{Background and Related Work}
\label{sec:background}
This section reviews the evolution of the \pol{} problem, from centralized solutions with trusted anchors, through partially distributed protocols, to fully decentralized and permissionless approaches. It situates this paper's contributions within those developments and identifies the key gaps our work aims to address.

\begin{figure}[!ht]
    \centering
    \includegraphics[width=\textwidth]{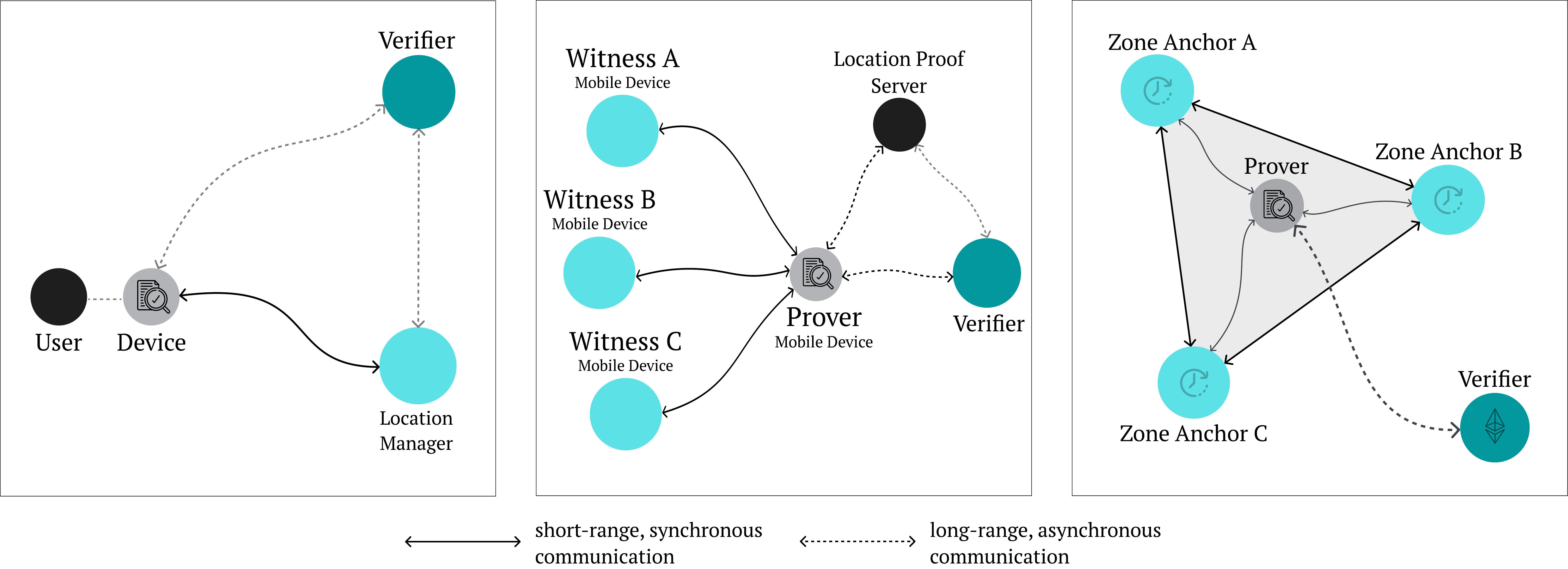}
    \caption{Evolution of \pol{} infrastructures from centralised to decentralized models. On the left, centralised systems rely on a trusted location manager to issue location proofs~\cite{waters2003secure}. In the middle, progressively distributed models involve peer-to-peer interactions between provers and witnesses~\cite{zhu2011applaus}. On the right, decentralized systems eliminate the need for a central authority, enabling direct peer-to-peer attestations and storage in a distributed ledger~\cite{foam-white-paper}.}
    \label{fig:related-work}
\end{figure}

\subsection{Trusted and Centralized Architectures}
\label{sec:related-work-trusted-centralized}

The notion of enabling \pol{}—that is, providing secure proofs that a device was present at a specific location—dates back to the early 2000s. A foundational contribution is the work by Waters and Felten~\cite{waters2003secure}, who proposed a proximity-based system with integrity and privacy guarantees. Their model, shown on the left of Figure~\ref{fig:related-work}, assumed a controlled setup with a verifier, a location manager, and a tamper-resistant device trusted by the verifier. Typical use cases included lending equipment (e.g., university computers) or house arrest monitoring, where the device must remain within physical boundaries. Although dependent on trusted wireless LAN infrastructure, this work introduced round-trip time and signal-propagation delays as countermeasures to proxy attacks.

Subsequent efforts, like Saroiu and Wolman~\cite{saroiu2009enabling}, further shaped the \pol{} concept and highlighted applications such as store rewards, location-gated content, and presence verification for voting. Graham and Gray~\cite{graham2009protecting} proposed SLVPGP, which shifted trust to tamper-resistant modules. VeriPlace~\cite{luo2010veriplace} expanded this with a privacy-aware architecture distributing trust across several entities, originally targeting platforms like Yelp. Later works streamlined proof generation via Wi-Fi infrastructure~\cite{javali2016alice} or proposed centralized yet formally secure geo-tamper-resistant designs~\cite{akand2021privacy}.

Together, these trusted and centralized systems showed that secure location proofs are feasible using short-range wireless signals and trusted anchors. However, their dependence on central authorities limits scalability and exposes them to collusion risks.

\subsection{Progression to Distributed Protocols}
\label{sec:related-work-distributed-decentralized}

To reduce reliance on fully trusted entities, more distributed \pol{} designs began to emerge. VeriPlace~\cite{luo2010veriplace} already hinted at a multi-entity architecture for privacy without dedicated tamper-resistant hardware, albeit with increased complexity.

APPLAUS~\cite{zhu2011applaus} was among the first explicitly distributed proposals, combining location privacy with Bluetooth-based local interactions. Provers collected signatures from nearby witnesses and stored the resulting proofs on an untrusted server (middle of Figure~\ref{fig:related-work}). While a Certificate Authority (CA) still issued identities, the location-proof server was distrusted. The work aimed at improving privacy and reducing single points of failure. STAMP~\cite{wang2016stamp} and PROPS~\cite{gambs2014props} added group signatures, but faced challenges when provers colluded with witnesses. SPARSE~\cite{nosouhi2018sparse} addressed this by removing the prover’s control over witness selection and leveraging crowd density for secure attestations.

Other approaches employed advanced cryptography to secure and anonymize claims. Dupin et al.~\cite{dupin2018location} used secure multi-party computation (MPC) to let multiple semi-honest entities jointly verify location, at the cost of significant computational overhead. These mid-stage protocols showed how trust could be distributed to reduce central dependencies, but often introduced new trade-offs in complexity, resource use, or environmental assumptions.

\subsection{Decentralized Systems}
\label{sec:related-work-fully-trustless}

With the rise of blockchain technology, decentralized and permissionless \pol{} architectures gained traction. Amoretti et al.\cite{amoretti2018blockchain} combined short-range wireless protocols with blockchain-backed storage to achieve tamper-evident and censorship-resistant proofs without relying on a trusted server. Nasrulin et al.\cite{nasrulin2018robust} built on this by formalizing spatio-temporal verification requirements and deploying a permissioned blockchain for supply-chain tracking.

Subsequent work layered privacy onto decentralized ledgers. Wu et al.\cite{wu2020blockchain} introduced Zero-Knowledge \pol{} (ZK-PoL), enabling presence proofs with minimal disclosure~\cite{bogdanov2025zero}. Nosouhi et al.\cite{nosouhi2020blockchain} added incentive mechanisms to reward honest witnesses, combining privacy with public blockchains. The FOAM protocol~\cite{foam-white-paper}, on the right of Figure~\ref{fig:related-work}, advanced zone-based time synchronization, coordinating radio beacons to achieve consensus on clock offsets without centralized control.

Complementary work by Pournaras~\cite{pournaras2020proof} explored decentralized civic engagement via Proof-of-Witness-Presence, where participants verify each other's presence in smart-city settings. Altogether, decentralized \pol{} systems now combine local distance bounding, digital signatures, consensus-based ledgers, zero-knowledge proofs, and incentive schemes, as systematized by Brito et al.~\cite{brito2025decentralized}.

In summary, \pol{} protocols have evolved from early centralized systems with trusted infrastructure, through distributed models mitigating single points of failure, to decentralized frameworks relying on cryptographic primitives and permissionless consensus. While each addresses different real-world scenarios—from retail and crowd-sensing to supply chain—there remains no unified framework for classifying these efforts by core \pol{} properties. Building on these developments, we propose a consolidated view of \pol{} across trust models and witness structures, aiming to guide practitioners in navigating design choices, assessing verifiability, and selecting appropriate enabling mechanisms.
% -- brief recap of decentralized PoL, existing classification limits

\section{A Taxonomy of Location-Proof Use Cases}
\label{sec:taxonomy}
In this section, we propose a structured taxonomy for \pol{} protocols and their use cases. We begin with core definitions before introducing four major domains of the taxonomy, which are summarized in Table~\ref{tab:taxonomy-overview}.

\subsection{Foundations and Basic Definitions}
\label{sec:taxonomy-foundations}

We adopt the standard \pol{} model with three primary roles~\cite{brito2025decentralized}:
\begin{itemize}
  \item \textbf{Prover} ($\mathcal{P}$): the party claiming a location statement, 
    e.g.\ ``I was in region~$R$ at time~$t$.'' 
  \item \textbf{Verifier} ($\mathcal{V}$): checks whether $\mathcal{P}$'s claim 
    is valid under the given threat model and trust assumptions.
  \item \textbf{Witnesses} ($\{\mathcal{W}_i\}$): optional external participants who measure or attest $\mathcal{P}$'s presence. They may be physical anchors (access points, radio beacons) or other peer devices in the environment.
\end{itemize}

A \emph{location proof} is a cryptographically bound statement ensuring the correctness of ``$\mathcal{P}$ was in $R$ at time $t$,'' optionally accompanied by measured distances, timestamps, or other evidence from the witnesses. Following classical interactive proof systems, we require:
\begin{itemize}
  \item \textbf{Soundness.} If the prover is not in $R$ at $t$, the verifier should reject any maliciously generated proof with overwhelming probability.
  \item \textbf{Completeness.} If the prover is actually in $R$ at $t$, and all parties, that is the prover and eventual witnesses, behave honestly, the proof should be accepted, with overwhelming probability, by an honest verifier.
\end{itemize}

For \pol{}, soundness can subdivide into proximity soundness (preventing distance fraud) and 
temporal soundness (ensuring correct time references), as elaborated in the following.

\begin{table}[ht]
\centering
\caption{Taxonomy Summary: Core Domains in \pol{} Systems}
\label{tab:taxonomy-overview}
\small
\renewcommand{\arraystretch}{1.1}
\begin{tabular}{p{3.6cm} p{8.4cm}}
\toprule
\textbf{Domain} & \textbf{Key Aspects} \\
\midrule
\textbf{(1) Cryptographic \newline Guarantees} & Soundness (spatial, temporal), authenticity, integrity, consensus, privacy \\
\textbf{(2) Spatio-temporal \newline Synchronization} & Source of spatial and time references, required precision, robustness to spoofing or drift \\
\textbf{(3) Trust and \newline Witness Model} & Prover-managed, verifier-controlled, partially distributed, or fully decentralized \\
\textbf{(4) Interaction and  \newline Overhead} & Interactive vs.\ non-interactive, proof size, verification cost, real-time vs.\ offline use \\
\bottomrule
\end{tabular}
\end{table}

\subsection{Domain 1: Cryptographic Guarantees}
\label{sec:taxonomy-crypto}

Under this domain, we collect the specialized security properties that refine completeness and soundness in \pol{} scenarios:

\begin{itemize}
  \item \textbf{Proximity Soundness (via Distance Bounding).} Constitutes a practical extension to localization techniques to prevent relay (``mafia'') attacks, ensuring a remote prover cannot pretend to be close simply by forwarding signals \cite{capkun2006secure,akand2023contribution}. Protocols employ tight round-trip timing or rapid bit exchanges to verify that $\mathcal{P}$ is within a certain distance.

  \item \textbf{Temporal Soundness (via Trusted Time Sources).} Ensures that the timestamp $t$ associated with a location proof reflects the actual moment of the event and cannot be falsified or replayed. This is typically achieved through trusted anchors or time-stamping authorities that bind a verifiable time reference to the proof. Without this guarantee, adversaries could backdate or reuse valid proofs to falsely claim presence at other times~\cite{nasrulin2018robust,foam-white-paper}.

  \item \textbf{Authenticity.} Only the legitimate prover and witnesses can generate or attest to location proofs in their own name (no spoofing under another’s credentials). This is typically enforced via public-key signatures or secure device identities, like keys backed by Trusted Platform Modules.

  \item \textbf{Consensus and Event Ordering.} In distributed or decentralized systems, additional guarantees are needed to ensure that multiple witnesses or nodes agree on the order and integrity of events across the system~\cite{brito2025decentralized}. This often requires consensus algorithms such as PBFT~\cite{castro1999practical} or blockchain-based ledgers~\cite{nasrulin2018robust}, which collectively maintain an append-only, tamper-evident log of location claims. These mechanisms complement temporal soundness by preventing reordering or deletion of proofs after the fact~\cite{amoretti2018blockchain,brito2025decentralized}.
    
  \item \textbf{Privacy.} Many use cases require minimal disclosure. For instance, $\mathcal{P}$ might reveal only that it was in a certain region without disclosing exact coordinates or identity. Zero-knowledge proofs~\cite{li2020privacy,bogdanov2025zero,wu2020blockchain} can enforce location membership while hiding extraneous details, albeit at greater computational cost.
\end{itemize}

In high-assurance scenarios, such as legal forensics, liability tracking, or e-government, additional properties like non-repudiation (ensuring the prover cannot deny a valid proof) and non-transferability (preventing reassignment of proofs to others) may be required. Achieving these guarantees in decentralized, privacy-preserving settings remains an open challenge. Future work may explore new digital signature schemes, identity-bound proofs, or co-signed attestations to address these needs.

\subsection{Domain 2: Spatio-temporal Synchronization}
\label{sec:taxonomy-spacetime}

Location proofs revolve around two fundamental aspects: where the prover is and when they are there. Systems thus differ not only in who provides the space and time references but also in how precise those references must be.

\subsubsection{Spatial Reference and Precision}

\begin{itemize}
\item \textbf{Reference Mechanism.} This describes how the prover’s physical location is determined, which varies across use cases:
\begin{itemize}
  \item \emph{Prover-located:} The prover self-reports (e.g.,\ via GPS). Simple to deploy, but prone to spoofing if not combined with secure hardware or distance bounding~\cite{waters2003secure}.
  \item \emph{Single anchor:} A location server or hardware beacon is controlled by the verifier. It is often used in centralized protocols where trust is placed in a physically secured anchor \cite{saroiu2009enabling,waters2003secure}.
  \item \emph{Multi-witness triangulation:} Multiple independent anchors or peer devices collectively bound the prover's position, characteristic of more fault-tolerant or decentralized systems~\cite{zhu2011applaus,nasrulin2018robust,foam-white-paper,brito2025decentralized}.
\end{itemize}

\item \textbf{Spatial Precision.}
This captures the granularity of the location claim, how accurately the system must localize the prover. Required precision can range from coarse (e.g., city- or region-level) to fine-grained (sub-meter), depending on the application's sensitivity and risk profile. Systems adopt different approaches based on these needs:
\begin{itemize}
  \item \emph{Coarse precision via GPS or single-anchor models:} Suitable for low-risk applications such as retail rewards or general mobility analytics, where city- or district-level location is sufficient~\cite{pournaras2020proof}.
  \item \emph{Short-range or specialized radio techniques:} Technologies like UWB or WiFi round-trip timing reduce localization error in room-scale or indoor environments. When combined with distance-bounding protocols, they also support security-critical applications that require proximity enforcement~\cite{alexandrusecure,akand2023contribution,capkun2006secure}.
  \item \emph{Dense witness coverage:} Multi-lateration using multiple vantage points or peer witnesses enables sub-meter multidimensional accuracy with fault-tolerance, often needed in critical infrastructure or industrial settings~\cite{chiang2009secure,foam-white-paper,brito2025decentralized}.
\end{itemize}

\end{itemize}

\subsubsection{Temporal Reference \& Precision}

\begin{itemize}
\item \textbf{Time Source.} This defines who is trusted to timestamp the prover’s location proof. The protocol must determine when the prover was at a location:
\begin{itemize}
  \item \emph{Prover’s local clock:} Minimal overhead, but untrusted if $\mathcal{P}$ 
  can alter timestamps for replay attacks.
  \item \emph{Verifier or Trusted Witness clock:} A central entity timestamps proofs; typical 
  in semi-centralized protocols \cite{waters2003secure}.
  \item \emph{Distributed ledger or witness-based clock:} 
  Achieves robust time integrity by ordering events via consensus~\cite{foam-white-paper,brito2025decentralized}. This suits high-collusion risk scenarios but raises the bandwidth and latency overheads.
\end{itemize}

\item \textbf{Temporal Precision.}
This captures how exact the timing information must be for the proof to be considered valid. Some applications demand only day-level granularity (e.g.,\ “present sometime on Monday”), while others require second- or even millisecond-level timestamps (e.g.,\ traffic management, real-time access control):
\begin{itemize}
  \item \emph{High-precision timing:} sub-second or sub-millisecond typically relies on synchronized protocols like NTP/PTP, or a blockchain with very tight block intervals~\cite{survey-dist-consensus}.
  \item \emph{Lower-precision needs:} can tolerate simpler clock models or occasional drift (e.g.,\ verifying presence within a 24-hour window)~\cite{foam-white-paper}.
  \item \emph{Replay mitigation:} If the adversary can shift $t$ by even a few seconds 
  to re-use an old proof, the system must maintain strict temporal soundness 
  (e.g.,\ \emph{time bounding} or short validity windows)~\cite{nasrulin2018robust}.
\end{itemize}

\end{itemize}

Taken together, the location reference mechanism (self-reported vs.\ anchored vs.\ multi-lateration) and the time source (local clock vs.\ anchor timestamp vs.\ consensus ordering) form the core of spatio-temporal synchronization. The chosen approach must align with each use case’s required precision, as well as its threat model (e.g.,\ do we fear adversaries forging sub-meter location or shifting timestamps by seconds?). As higher fidelity in space-time typically demands more infrastructure or more complex protocols, system designers must eventually balance security needs against deployment overhead.

\subsection{Domain 3: Trust and Witness Models}
\label{sec:taxonomy-trust}

This domain captures who is trusted to observe, validate, and attest to the prover’s location. It reflects the underlying trust assumptions of the system, whether centralized or distributed, and how witnesses are selected, authenticated, or incentivized. The model also affects the system's resilience to collusion, fault tolerance, and scalability:

\begin{itemize}
  \item \textbf{Prover-Managed.} The prover $\mathcal{P}$ is assumed to be trustworthy or equipped with tamper-resistant hardware, allowing it to self-certify its location. This approach has minimal overhead and works well in controlled environments (e.g., corporate devices), but offers limited resistance to spoofing or collusion~\cite{luo2010veriplace,javali2016alice}.

  \item \textbf{Verifier-Managed.} A location anchor or server is controlled directly by the verifier $\mathcal{V}$, who certifies the prover's presence. This model dominates early \pol{} systems~\cite{waters2003secure} and simplifies deployment, but introduces a central point of failure and requires complete trust in the verifier.

  \item \textbf{Partially Distributed.} A limited number of known entities (e.g., location anchors, authorized witnesses) jointly verify location claims. These may be coordinated through a certificate authority or pre-established trust relationships. This model balances decentralization and manageability, and is typical of mid-scale protocols with privacy or availability goals~\cite{zhu2011applaus}.

  \item \textbf{Fully Decentralized or Permissionless.} Witnessing is performed by an open and untrusted set of peers. Verifiability arises from quorum-based or consensus mechanisms (e.g., threshold cryptography, BFT agreement), and often includes incentive layers such as staking, token rewards, or smart contracts to promote honest behavior~\cite{nasrulin2018robust,foam-white-paper,nosouhi2020blockchain,brito2025decentralized}. This model enables scalability and strong collusion resistance, but also introduces coordination and latency overheads.
\end{itemize}

\subsection{Domain 4: Interaction Model and Overhead}
\label{sec:taxonomy-interaction}

This domain describes how location proofs are constructed and verified—whether they require real-time interaction or can be performed offline, and what computational or bandwidth costs are involved. These choices impact system usability, deployability in constrained settings, and the level of guarantees (e.g., proximity soundness, auditability, scalability):

\begin{itemize}
  \item \textbf{Interactive Proofs (Challenge–Response).}
  The prover must engage in a time-sensitive exchange with a verifier or witness (e.g., bit-level distance bounding). These are essential in contexts that demand secure proximity guarantees or live validation~\cite{capkun2006secure,akand2023contribution}.

  \item \textbf{Non-Interactive (Offline) Proofs.}
  The prover collects signed attestations or zero-knowledge statements from witnesses and later presents them to the verifier without needing further communication. This model is ideal for asynchronous auditing (e.g., supply chains, mobility attestations), but offers weaker proximity guarantees~\cite{foam-white-paper,nosouhi2020blockchain,nasrulin2018robust,amoretti2018blockchain}.

  \item \textbf{Proof Size and Verification Complexity.}
  Some zero-knowledge-based \pol{} protocols~\cite{li2020privacy,bogdanov2025zero,wu2020blockchain} produce large, computation-heavy proofs, especially when encoding distance or privacy constraints. In contrast, lightweight signature-based schemes offer lower overhead but less expressive power and weaker security guarantees~\cite{amoretti2018blockchain}.
\end{itemize}

\subsection{Summary and Discussion}
\label{sec:taxonomy-summary}

Table~\ref{tab:taxonomy-overview} summarizes the domains for classifying \pol{} systems: cryptographic guarantees, spatio-temporal synchronization, trust and witness models, and interaction or overhead. Each reflects a distinct design concern, from what the proof must guarantee to how location and time are established and verified.

This taxonomy aims to help designers navigate trade-offs between verifiability, privacy, and infrastructure complexity, enabling protocol comparisons across security models and better alignment with application needs.

Nonetheless, the taxonomy abstracts away some real-world complexity. Hybrid architectures that blend interaction models or trust configurations may defy strict categorization. Emerging deployments may feature dynamic witnesses, layered identity systems, or continuous sensing, all of which challenge these static models~\cite{brito2025decentralized}. New adversarial threats, such as AI-assisted spoofing or environmental emulation, may also shift current assumptions.

Future refinements could extend the taxonomy to address these hybrid and dynamic settings, or incorporate additional dimensions such as usability, legal constraints, or application-specific needs. In the next section, we introduce a methodology for selecting appropriate protocol designs by mapping requirements onto the proposed domains.

% -- classification dimensions, table of scenarios

\section{Methodology for Matching Use Cases to PoL Requirements}
\label{sec:methodology}
Having established our taxonomy in Section~\ref{sec:taxonomy}, we now present a five-step methodology for translating application-specific requirements into a concrete \pol{} protocol design. Each step aligns with one or more domains of the taxonomy and culminates in a set of cryptographic and infrastructural decisions.

\subsection{Step 1: Profile the Use Case}
\label{sec:methodology-step1}

The process begins by outlining the application’s operational and security requirements across four axes. This step yields a structured requirement profile that will guide further design decisions.

\begin{itemize}
    \item \textbf{Spatial and Temporal Precision.}
    Define the required granularity of the location proof (e.g., sub-meter, room-level, city-level) and the acceptable time window or resolution (seconds, minutes, or daily presence).

    \item \textbf{Threat Model.}
    Characterize expected adversaries: are relay attacks, timestamp tampering, or witness collusion realistic threats? What forms of spoofing must the system withstand?

    \item \textbf{Privacy and Disclosure.}
    Determine the level of information that must be protected. Should the proof reveal exact coordinates, or just region membership? Are there legal or regulatory constraints on disclosure?

    \item \textbf{Deployment Environment.}
    Identify device and network constraints. Are devices battery-constrained or capable of interactive sessions? Is the network intermittent? Is the system real-time or audit-based?
\end{itemize}

\subsection{Step 2: Classify the Use Case in the Taxonomy}
\label{sec:methodology-step2}

Next, the profile is mapped onto the four taxonomy domains from Table~\ref{tab:taxonomy-overview}:

\begin{itemize}
    \item \textbf{Cryptographic Guarantees.}
    Identify the required properties: spatio-temporal soundness, authenticity, integrity, consensus, or privacy preservation.

    \item \textbf{Spatio-temporal Synchronization.}
    Assess how precise and reliable the system’s spatial and temporal references must be. Choose between GPS, verifier-managed anchors, or distributed consensus.

    \item \textbf{Trust and Witness Model.}
    Choose a trust architecture: centralized (verifier-managed), partially distributed (registered anchors), or decentralized (open witnesses with incentives or quorum).

    \item \textbf{Interaction and Overhead.}
    Establish the acceptable interaction mode and cost model. Is real-time challenge–response feasible? Can the system tolerate large proof sizes or limited connectivity?
\end{itemize}

This mapping narrows the solution space by identifying which protocol properties are non-negotiable.

\subsection{Step 3: Select Cryptographic Building Blocks}
\label{sec:methodology-step3}

With the taxonomy placement in hand, designers can select appropriate cryptographic tools:

\begin{itemize}
    \item \textbf{Distance Bounding}~\cite{capkun2006secure,akand2023contribution,brito2025decentralized}. Needed for high-precision or proximity-enforcing use cases.

    \item \textbf{Time Anchoring}~\cite{nasrulin2018robust,brito2025decentralized}.
    Achievable through trusted timestamps or consensus-based ordering on a ledger.

    \item \textbf{Standard Authenticity and Integrity Mechanisms}~\cite{brito2025decentralized}.
    Public-key signatures, secure enclaves, or hash-chained logs ensure integrity and provenance. Threshold or Group Signatures~\cite{brandao2019threshold,brito2025decentralized} are useful for schemes requiring multiple witnesses to jointly attest a statement.

    \item \textbf{Zero-Knowledge Proofs}~\cite{li2020privacy,wu2020blockchain}.
    Used to prove location assertions without revealing fine-grained coordinates or prover identity.
\end{itemize}

When resource limitations (e.g., mobile environments) conflict with the above requirements, fallback strategies must be considered—e.g., reducing proof frequency or omitting distance bounding.

\subsection{Step 4: Choose Trust and Witness Infrastructure Models}
\label{sec:methodology-step4}

The trust and witness infrastructure models determine how the protocol operates at the system level:

\begin{itemize}
    \item \textbf{Single Anchor (Centralized)}~\cite{waters2003secure,akand2023contribution}.
    Suitable for low-risk applications with stable infrastructure and strong control over anchor security, but limited in coverage or fault tolerance.

    \item \textbf{Partially Distributed}~\cite{zhu2011applaus,luo2010veriplace}.
    Combines a small set of trusted anchors to improve coverage and resilience, without full decentralization.

    \item \textbf{Fully Decentralized}~\cite{brito2025decentralized,foam-white-paper}.
    Witnessing is open to untrusted participants, relying on threshold consensus or blockchains to enforce integrity. Highly robust against collusion, though more complex to operate and often paired with economic incentives.
\end{itemize}

The number, coverage, and mobility of anchors must align with the desired spatial precision. Higher granularity generally implies denser and better-placed infrastructure.

\subsection{Step 5: Define the Interaction Model and Estimate Overhead}
\label{sec:methodology-step5}

Finally, the designer specifies how location proofs are produced and verified in practice:

\begin{itemize}
    \item \textbf{Interactive Protocols.}
    Required for real-time checks or distance bounding, but involve latency-sensitive communication and tighter hardware timing constraints.

    \item \textbf{Non-Interactive (Offline) Proofs.}
    Preferred for post-event validation or intermittent connectivity scenarios, using signed statements or zero-knowledge proofs.

    \item \textbf{Cost and Efficiency.}
    Analyze whether proof sizes, computation, and bandwidth demands match the device and application context, especially for IoT, edge, or mobile deployments.
\end{itemize}

This step ensures that the protocol is not only secure, but feasible to run in the intended setting.

\subsection{Discussion and Trade-Offs}
\label{sec:methodology-discussion}

This methodology shows a structured path from application-level requirements to a tailored \pol{} design, grounded in the four taxonomy domains introduced in Section~\ref{sec:taxonomy}. It enables reasoning about trade-offs between verifiability, privacy, interactivity, and deployment complexity.

For instance, a transit ticketing system needing second-level timestamping and moderate spatial accuracy might rely on a short-range anchor and offline proof submission. In contrast, high-assurance identity verification may call for interactive distance bounding, quorum-based witness validation, and zero-knowledge disclosures.

Nonetheless, this methodology abstracts away several real-world challenges. It assumes that application requirements can be cleanly mapped to static design dimensions, whereas in practice, use cases may evolve over time or operate across multiple modes (e.g., interactive during onboarding, offline during regular use). Hybrid deployments that mix centralized and decentralized witnesses, or dynamically switch between precision levels, are difficult to model with the current framework. Moreover, the methodology does not yet account for economic incentives, user experience constraints, or regulatory compliance workflows, factors that increasingly shape system adoption in practice.

Future work could enrich this methodology with formal decision models or tool-assisted workflows to guide protocol selection. It may also benefit from empirical validation across real deployments, and from integration with broader system concerns such as identity management, data life cycle control, and composability with other attestations (e.g., proof of identity, proof of intent). As the \pol{} design space expands, a modular, extensible framework for reasoning about trust, precision, and accountability across heterogeneous environments will become increasingly necessary.

% -- systematic process, step-by-step approach

\section{Illustrative Use Cases}
\label{sec:usecases}
A diverse range of real-world applications benefit from \pol{} systems, reflecting the growing need for verifiable presence data in modern societies~\cite{brito2025decentralized}. Among the most prominent \pol{} use-case domains are:

\begin{itemize}
    \item \textbf{Civic and Governance:} Voter presence verification at polling stations, smart-city crowd-sensing for public resource allocation, or location-based eligibility checks (e.g., verifying residency for local benefits)~\cite{pournaras2020proof}.

    \item \textbf{Supply Chain and Logistics:} 
    Authenticating product handoffs along distribution chains (to thwart counterfeits in pharmaceuticals or high-value goods), and ensuring goods remain in authorized areas~\cite{nosouhi2020blockchain,nasrulin2018robust}.

    \item \textbf{Delivery and Transport Services:} Proof of package handover (reducing ``lost shipment'' disputes), ride-sharing liability checks, or verifying IoT sensor deployments~\cite{amoretti2018blockchain}.

    \item \textbf{Retail and Entertainment:}
    Issuing location-based promotions, tickets, or exclusive content to users physically visiting a store, stadium, or event~\cite{saroiu2009enabling,luo2010veriplace}.

    \item \textbf{Content Authentication:} 
    Demonstrating that digital media (photos, videos) were captured 
    at a certain place and time, mitigating deepfake concerns \cite{brito2025decentralized}.
\end{itemize}

Collectively, these scenarios span everything from high-stakes (e.g., e-voting with strong adversary models) to low-risk or coarse-grained contexts (e.g., city-level coupons). Some demand sub-meter precision and near-real-time assurance, while others settle for ex-post auditing of day-level location logs. Certain applications must also preserve privacy via selective disclosure or zero-knowledge proofs, whereas others are content with public logs of presence.

To illustrate how our five-step methodology (Section~\ref{sec:methodology}) applies to different constraints, we now explore three specific examples: \emph{(1) Retail E-Coupon Rewards}, \emph{(2) E-Voting with Strict Proximity}, and \emph{(3) Supply Chain Auditing with Medium Precision}. Each scenario shows a combination of spatial precision, time constraints, threat model, and privacy or resource demands, leading to distinct protocol configurations.

\subsection{Use Case 1: Retail E-Coupon Rewards}
\label{sec:usecases-retail}

\textbf{(1) Requirements.} A store issues digital coupons only to customers physically present on-site~\cite{saroiu2009enabling,luo2010veriplace}. Spatial precision is coarse (100–200m), with same-day temporal granularity. Threats are mild (little collusion or relay risk), and privacy demands are low, as the location claim is coarse, voluntary, and not linked to sensitive personal data.

\noindent
\textbf{(2) Taxonomy Mapping.}
\begin{itemize}
\item \emph{Guarantees}: Basic authenticity and integrity; no proximity or time bounding.
\item \emph{Synchronization}: Single anchor (e.g., Wi-Fi beacon), day-level timestamp.
\item \emph{Trust Model}: Centralized (store-managed).
\item \emph{Interaction and Overhead}: Non-interactive; low computation and communication.
\end{itemize}

\noindent
\textbf{(3) Cryptographic Building Blocks.} A public-key digital signature scheme suffices. No distance bounding or zero-knowledge is needed.

\noindent
\textbf{(4) Trust and Witness Infrastructure.} The store acts as a trusted anchor. A BLE or Wi-Fi beacon issues signed proofs upon customer interaction.

\noindent
\textbf{(5) Interaction and Overhead.} Proofs are collected offline and shown during checkout. System demands are low, with minimal real-time communication.

\subsection{Use Case 2: Supply Chain Auditing}
\label{sec:usecases-supplychain}

\textbf{(1) Requirements.} A logistics provider tracks goods across distribution centers~\cite{amoretti2018blockchain,nasrulin2018robust,nosouhi2020blockchain}. Required precision is moderate (\textpm{}50 meters); timestamps are needed at hour or day-level granularity. Threats include tampering and forgery, but not advanced relays. Privacy is moderate: visibility may be limited to stakeholders.

\noindent
\textbf{(2) Taxonomy Mapping.}
\begin{itemize}
\item \emph{Cryptographic Guarantees}: Authenticity and tamper-evidence; no sub-meter proximity.
\item \emph{Spatio-temporal Synchronization}: Anchors provide signed location and time.
\item \emph{Trust Model}: Partially distributed (e.g., partner-operated checkpoints).
\item \emph{Interaction and Overhead}: Non-interactive; ex-post verifiability.
\end{itemize}

\noindent
\textbf{(3) Cryptographic Building Blocks.} Anchor-issued digital signatures provide basic presence proofs. Ledger-anchored hash chains are used to ensure tamper-evidence and enable aggregation.

\noindent
\textbf{(4) Trust and Witness Infrastructure.} Each checkpoint hosts one or more trusted anchors, ideally run by different partner firms. These anchors issue signed location-time records, minimizing reliance on a single entity~\cite{nasrulin2018robust}.

\noindent
\textbf{(5) Interaction and Overhead.} Devices collect signed proofs during handovers; verification is deferred to a later auditing phase. Overheads are modest, matching the use case's moderate assurance requirements.

\subsection{Use Case 3: E-Voting with Location Binding}
\label{sec:usecases-evoting}

\textbf{(1) Requirements.} An election authority wants to ensure that votes are only cast by citizens physically present at designated polling stations during election day~\cite{pournaras2020proof,li2020privacy}. Spatial precision is fine-grained (under 10 meters) to prevent location spoofing. Temporal precision must support real-time validation to prevent multiple voting. The threat model is strong: relay and collusion attacks are realistic. Privacy is paramount to protect voter anonymity and external location leakage.

\noindent
\textbf{(2) Taxonomy Mapping.}
\begin{itemize}
\item \emph{Cryptographic Guarantees}: Proximity soundness (distance bounding), time integrity, and privacy via selective disclosure.
\item \emph{Spatio-temporal Synchronization}: Multi-witness triangulation and synchronized clocks; possibly combined with tamper-evident logs or consensus.
\item \emph{Trust Model}: Partially distributed (independent anchors) or fully decentralized (to avoid centralized abuse).
\item \emph{Interaction and Overhead}: Interactive challenge-response; high computational and coordination complexity.
\end{itemize}

\noindent
\textbf{(3) Cryptographic Building Blocks.} Distance bounding protocols~\cite{brito2025decentralized} are used for real-time proximity soundness. A permissioned ledger or consensus system (e.g., PBFT) enforces temporal ordering and protects against timestamp forgery~\cite{nasrulin2018robust,brito2025decentralized}. Zero-knowledge proofs~\cite{wu2020blockchain} support privacy-preserving presence claims without revealing sensitive metadata.

\noindent
\textbf{(4) Trust and Witness Infrastructure.} Polling locations are equipped with multiple independent anchors operated by separate trusted entities (e.g., civic oversight bodies or parties). These anchors jointly participate in the proof process and sign ephemeral measurements~\cite{brito2025decentralized}.

\noindent
\textbf{(5) Interaction and Overhead.} Voters perform real-time distance-bounding with nearby anchors, which may require specialized radio protocols~\cite{brito2025decentralized}. Collected measurements are combined and recorded in a ledger for auditability. Proofs are privacy-preserving but computationally heavier. The overhead is justified by the high-stakes and high-threat model of elections.

\subsection{Summary and Comparison}
\label{sec:usecases-comparison}

Table~\ref{tab:usecases-comparison} summarizes the design choices for the three scenarios analyzed. These examples demonstrate how varying combinations of spatial precision, timing constraints, privacy demands, and adversarial strength lead to distinct \pol{} protocol configurations.

\begin{table}[ht!]
    \centering
    \caption{Comparison of illustrative \pol{} scenarios.}
    \label{tab:usecases-comparison}
    \small
    \renewcommand{\arraystretch}{1.1}
    \begin{tabular}{p{2.3cm}p{1.8cm}p{1.3cm}p{2.1cm}p{2.4cm}}
    \toprule
    \textbf{Use Case} & \textbf{Precision} & \textbf{Threat Model} & \textbf{Trust Model} & \textbf{Proof Style} \\
    \midrule
    Retail Coupons & Coarse \newline (100--200m) & Low & Centralized & Non-Interactive \\
    Supply Chain & Medium \newline (\textpm{}50m) & Medium & Partially \newline Distributed & Non-Interactive \\
    E-Voting & Fine \newline (\textless10m) & High & Partially/Full \newline Decentralized & Interactive \\
    \bottomrule
    \end{tabular}
\end{table}

% -- short examples, walk-through of the methodology/logic

%\section{Formalizing Semantics with Logic}
%\label{sec:logic}
%\input{sections/06-logic}
% -- define predicates, modal/temporal operators, proofs

%\section{Discussion and Future Directions}
%\label{sec:discussion}
%\input{sections/07-discussion}
% -- opportunities, research challenges, expansions

\section{Conclusion and Future Work}
\label{sec:conclusion}
We introduced a structured taxonomy and methodology for designing Proof-of-Location (\pol{}) systems, spanning four core domains: cryptographic guarantees, spatio-temporal synchronization, trust models, and interaction and overhead. A five-step process maps real-world constraints to concrete protocol features, illustrated through three scenarios—retail coupons, supply chain auditing, and e-voting—each with distinct requirements in precision, trust, and privacy.

Looking forward, challenges remain in scaling \pol{} systems to real deployments, especially those involving decentralized ledgers, distance bounding, or zero-knowledge proofs. Balancing privacy with efficiency, modeling stronger adversaries, and standardizing verifiable location claims are promising future directions. As trust in location claims becomes as essential as identity, we envision \pol{} protocols becoming foundational tools across emerging digital societies.

%This paper has introduced a taxonomic view of Proof-of-Location (\pol{}) systems, encompassing dimensions such as cryptographic guarantees, spatio-temporal synchronization, trust and witness models, and protocol interaction overhead. Building on this taxonomy, we presented a methodology that guides practitioners from high-level use-case constraints (e.g., spatial precision, threat model, privacy) to appropriate protocol choices. To illustrate the approach, we walked through representative use cases, ranging from retail e-coupons to supply chain auditing and e-voting, highlighting how each scenario’s unique demands shape the selection of anchors, distance bounding, ledger-based timestamps, and more.
 
%Looking ahead, continued progress in standardizing PoL data formats, optimizing zero-knowledge protocols, and addressing sophisticated adversarial models will drive the further evolution of this field. Ultimately, we envision a future where robust, privacy-preserving \pol{} capabilities are as ubiquitous as digital signatures in online transactions, enabling trustworthy location claims across a broad spectrum of applications.
% -- final summary, key takeaways

% Paste BibTeX references here from .bbl file.
% Example:
% \bibliographystyle{plain}
% \input{nature.bbl}

\bibliographystyle{splncs04}
\bibliography{ref}

\end{document}